# The Role of the XBRL Standard in Optimizing the Financial Reporting

Grosu V., Hlaciuc E., Iancu E.,Petris R., Socoliuc M.

**Abstract**— When the financial information is difficult to produce, interpret, compare and analyze, we are put in the situation to face inconvenient consequences with negative repercussions, such as: the investor can give up the investment (with negative consequences on the risk equity market), the banks may not give loans, an auditor may not consider the financial statements as being credible etc.

These facts allow the introduction of this paper's main objective, the eXtensible Business Reporting Language (XBRL) which is an open standard, independent and international for the treatments, opportunity, correctness, efficiency and minor costs of the financial and economical information. The XBRL will be analyzed in the second part of the paper, the history of this electronic communication language will be described, as there will also be described the promoting organizations, the base technology (the WEB and XML architecture which will be the next stage of the internet programming), and the role it has within the chain of reporting between the XBRL consortium and the international accounting organizations IASB-CI. This taxonomy serves clearly every accounting and extra- accounting information made by the company. This information which is treated in present by resorting to various formats or structures (most times incompatible between them and the owners) will be standardized with the XBRL.

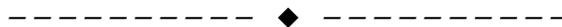

## 1 INTRODUCTION

*Short history regarding the actual IAS/ IFRS configuration*

According to the economical theory, the demand for accounting information comes from the economical asymmetry and the way the existing conflicts between the entity's management and the extern investors are solved (Hedy and Palepu, 2001). The financial statements and their reporting are placed among the most important available to the board of the company to communicate with the investors. On the developed financial markets, in order to consolidate the credibility of the information supplied by the managers third parties are implied, such as: legislative organizations, the Setters standards, auditors, information intermediaries (financial analysts and the rating companies).

Nevertheless, the credibility of the supplied information is not enough, these having also to be relevant, that is to be useful to the investor in making the decision to invest, or in choosing the portfolio structure. One of its objectives is to show that the publishing of the financial statements can offer new relevant information to the investors. The quality characteristics of the financial and economical information vary with the characteristics of the entities and the origin country (Collins and Kothari 1989, Alford and coat. 1993, Ball and coat. 2000). From the second half of the 80, among the science people of the financial and economical environment and its operators, has been spread the feeling of dissatisfaction regarding the capacity of the existent accounting policies and of the traditional accounting criteria to supply relevant information, not surprisingly being the fact that the accounting systems have evolved in parallel with the development of the economical systems. Several studies have shown a decline in the capacity of the profit or other accounting jobs in explaining the economical performance of the entity. The analysis of Chang (1999) Leo and Zarawin (1999), all reported to the economical and financial reality of the USA, show that the relation between the shares' value and profit, as that of the shares' prices, profit and "book values" has wakened in time. Partial results were also obtained by Francis and Schipper (1999) which, one on hand, tried to study the loss of profit relevance in explaining the shareholders' income, and on the other side sustain that the explicative power of the "book value" compared to the market value has not suffered deteriorations along the studied period (1952- 1994), but in contrary it seems to have raised. The mentioned researches belong to the economical literature domain which studies "the value of the financial and economical information relevance"- contained in the financial statements and, in general, the relations between the capital market and the accounting information.

The mentioned papers practically aim to determine the content of the information that results of the

- Grosu V. . *is from University of Suceava, 13 University Str., Suceava, Romania*
- Hlaciuc E. . *is from University of Suceava, 13 University Str., Suceava, Romania*
- Iancu E. *is from University of Suceava, 13 University Str., Suceava, Romania*
- Petris R. *is from the University of Suceava, 13 University Str., Suceava, Romania*
- Socoliuc M. *is from University of Suceava, 13 University Str., Suceava, Romania*

accounting data, so observing the variables as: share price variation, price volatility variation and transacted volume variation, in a period of time, centered around the data at which the information is published for the for the first time. The result of these tests lead to checking the efficiency degree of the information on the financial market, on a perfect market, the prices are adjusted instantly and completely as the new information appears, so that the market values are influenced by the data releases only on a short period of time. The first contribution in the domain is attributed, not randomly, to Fama and coauthors (1969), which analyze the adjustment process of the shares' prices tightly connected to the spreading of the new information. Fama's work is also important from a methodological point of view, because it establishes the bases of the analysis called "case study". These analyses and studies are destined to test the significant statistic reactions, on the financial markets, and also signal the appearance of any event that may influence the market value of a company. In general terms, the case studies are an important investigation instrument in analyzing the efficiency of the financial markets, the base idea which highlights this analysis type is that persistent and systematic income are not compatible to the hypothesis that the markets are efficient from the management information point of view, regarding to the behavior of the managers and shareholders. The special literature regarding case studies is impressive, under the methodological points of view but also the empirical one. In what regards the methodology, the researches made on this theme are very advanced, so that for the economical literature can be easily affirmed that "knowing *how* to do is known for a long time - and not how *not* to do- a case study" (Kothari and Warner, 2005). Relevant methodological contributions are also made by Peterson (1989), Armitage (1995) and McWilliams and Siegel (1997), while the empirical studies were mainly focused on cross-section type investigations and historical analysis.

In the cross-section studies, the objective of the research is almost always to analyze if the expected value, afferent to a spread of the income obtained over the entity's expectation (income that do not comply with the ones that should result from the market strengths' functioning), is different to the earning distribution stipulated by a theoretical model and nevertheless to verify of the empirical and theoretical distributions present different data compared to the initial statistics. The literature part and the case studies that connect directly to the present paper refer to the evaluation of the accounting policies changes consequences. In this regard, a lot of the research theme focus their attention to the motives for which changes should be made for the existing regulations, more than the "previous" and "after" analysis of these changes' consequences. On the other hand, the development of the financial markets and the increase of the entities' organizational structures complexity make more and more necessary the existence of studies that are capable to evaluate the effect of the changes in the data revealing systems regarding the entity's management activity. From this point of view, adopting the new IFRS by all the listed entities, in a series of countries, certainly is a great opportunity to make a case study whose relevance is to lead also to considerations that the present researches, regarding the accounting policies modifications, could not have studied a more important phenomenon such as this one is. In the existent literature, we can find studies (Ball 1972, Drahan 1993) that examine the individual effects, changes of principles or criteria such as modifications in the stock evaluation methods, changes in the computing methods of the debts to the employees or to the depreciation methods. The investors' sensibility analysis and of various accounting treatments were developed by Bartov and coauthors (2002) regarding German companies, by Anuer (1996) and Caramanolis- Cotelli and coauthors (1999) regarding the Swiss companies; in the mentioned countries, it is already possible to observe the entities that make financial statements according to various standards: national standards or accounting norms, European directives or the IAS/IFRS. The results of the researches obtained by present times show a greater relevance, to the investors, for the information made under the IAS/IFRS compared to the national accounting standards.

- The previous results are complying with the EU premises that, by adopting the IAS/IFRS, highlight the promoting of an explicit system, focused on the investors' information needs, according to the USA model and the Anglo-Saxon one. There are, still, some aspects of the IAS/IFRS that continue to be the subject of an intense debate (and critic) on behalf of the professionals, both in Europe and in the USA. Doubts regarding the Continental Europe economical entities' capacity to quickly adapt to the market demands were detailed by numerous authors, some of them foreseeing, at least for a short term, a disorientation period and even a diminution of transparency and result comparison, until the companies will have passed the initial phase (Sittle, 2004). Focusing on the IFRS, whose implementation and appliance have raised the biggest difficulties, we will especially study the IAS32, IAS39, IFRS4, IAS41 and the IFRS3. Regarding the IAS32 and the IAS39, the newest information refer to the notion of "fair value" for all the financial instruments, but also the remarkable expansion of the evaluation criterion for the fair value of the assets and liabilities of the financial statements. These changes or modifications were preceded and accompanied by an intense debate regarding the relevance of the information supplied by the fair value criterion, compared to the alternative criterion of the historical cost and the problems these two different evaluation methods bring, regarding the reliability of the evaluation's result (trade-off between relevance and reliability). Studies regarding the opportunity of the fair value are pretty numerous, we will only quote from the contributions of Barley and Hadadd (2003) and Chotorou, for which we can strongly affirm that represent a "masterwork" of this subject. Empirically speaking, the greatest part of the existent research on this theme refer to the USA experience, because in the economical and financial reality of the USA, the accounting norms have anticipated a few years ago the rules introduced by the IAS (SFAS107 and 115 regarding information rights and SFAS133 regarding the evaluation criteria), but these studies do not reach the same conclusions (Barth, 1993 and Barth and coauthors 1996) getting results in favor of the "relevance value" of the fair value criterion instead of the historical cost's. A different aspect have the conclusions of Nelson (1996), which state that the information on the "fair value" supplied by the appliance of the SFAS107 do not have a superior informing capacity compared to the input values in explaining the values assumed by the share prices. More convincing are the conclusions of Eccher and coauthors (1996), according to which the fair value of the portfolio titles has a significant capacity to explain the values observed by the share prices, whereas between the fair value of other assets and liabilities and the market value of the equity there would be no significant statistic relation. In a more recent study made by Khurana and Kim (2003), there has been found no proof to highlight or demonstrate a greater capacity of the fair value compared to the log value in explaining the market values, the study being made on a bigger sample of banks. The research activity dedicated to the problems brought by the actual IAS/IFRS configuration is relatively underdeveloped, (the up-to-date degree of the theme must be taken into consideration)- Gabriel Hernandez (2003) criticizing the structure of the SFAS133, and especially the accounting norms regarding hedge accounting (which are almost similar to the IAS39). These authors manifest their discontent to the norms that regulate hedge accounting in the IAS39, fact proven by numerous professional associations. Another aspect worthy of highlighting is the great complexity of the IAS39, especially in what regards the accounting treatment of the derived financial instruments (including the ones incorporated in other financial assets). The USA experience offers significant warnings, worthy to take into consideration, once with the appliance of the SFAS133 in December 2004, the FASB having to confront more than 183 "implementation problems".

**XBRL- the new economical- financial communication standard**

XBRL (eXtensible Business Reporting Language) is an open data standard for financial reporting. XBRL allows information modeling and the expression of semantic meaning commonly required in business reporting. XBRL is XML-based. It uses the XML syntax and related XML technologies such as XML Schema, XLink, XPath, and Namespaces to articulate this semantic meaning. One use of XBRL is to define and exchange financial information, such as a financial statement. The XBRL Specification is developed and published by XBRL International, Inc. (XII). XBRL is a standards-based way to communicate business and financial information. These communications are defined by metadata set out in taxonomies. Taxonomies capture the definition of individual reporting concepts as well as the relationships between concepts and other semantic meaning.

XBRLS is a simplified application profile of this standard intended to enable the non-XBRL expert to create both XBRL metadata and XBRL reports in a simple and convenient manner. At the same time, it seeks to improve the usability of XBRL, the interoperability among XBRL-based solutions and to reduce software development costs.

A wiki repository of XBRL projects is available to be freely explored and updated.

In typical usage, XBRL consists of an instance document, containing primarily the business facts being reported, and a collection of taxonomies (called a Discoverable Taxonomy Set (DTS)), which define metadata about these facts, such as what the facts mean and how they relate to one another. XBRL uses XML Schema, XLink, and XPointer standards.

The instance document begins with the <xbrl> root element. There may be more than one XBRL instance embedded in a larger XML document. The XBRL instance document itself holds the following information:

- Business Facts – facts can be divided into two categories
    o Items are facts holding a single value. They are represented by a single XML element with the value as its content.
    o Tuples are facts holding multiple values. They are represented by a single XML element containing nested Items or Tuples.

**In the design of XBRL, all Item facts must be assigned a context.**

- Contexts define the entity (e.g. company or individual) to which the fact applies, the period of time the fact is relevant, and an optional scenario. Date and time information appearing in the period element must conform to ISO 8601. Scenarios provide further contextual information about the facts, such as whether the business values reported are actual, projected, or budgeted.
- Units define the units used by numeric or fractional facts within the document, such as USD, shares. XBRL allows more complex units to be defined if necessary. Facts of a monetary nature must use a unit from the ISO 4217 namespace.
- Footnotes use XLink to associate one or more facts with some content.
- References to taxonomies, typically through schema references. It is also possible to link directly to a linkbase.

In Romania, the financial statements are made on paper forms, using MS programs which don't respect an electronic format standard, are put on flexible disks- although there aren't made laptops or desktops with reading/ writing units for these disks, and use morally- used data bases (DBF).

Romania's join to the EU makes necessary adopting a common language of financial-economical reporting, the XBRL (eXtended Business Reporting Language).

The financial statements, balance sheets and profit and loss account must be formatted according to a world standard. Therefore, each country elaborates its own taxonomy respecting the international standards.

Adopting the XBRL in Romania should drastically reduce costs of collecting and validating information from reporting, converting, disseminating and exchanging data.

Among the main beneficiaries of the XBRL adoption are: the Romanian Government, the Ministry of Finances, Intermediate agencies and organization of European funds' management, the Romanian Organization of Accounting experts and Authorized Accountants, the Romanian Chamber of Auditors, the local Romanian financial administration, the financial banking system and the Romanian commercial companies.

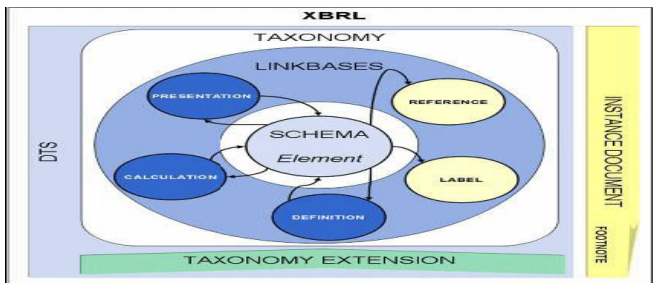

*Source:*
*http://www.iasb.org/NR/rdonlyres/95D8BAE9-24A1-40D1-A6B1-46F1B9F32436/0/IFRSTaxonomyGuide100080828.pdf*

**How does the XBRL confront the financial reporting probelms?**

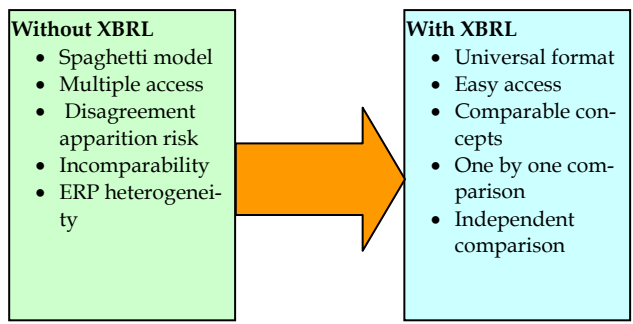

- XBRL needs no license;
- XBRL International is neuter and independent;
- the greatest part of the ERP is already malleable ;
- XBRL is adopted and implemented in the whole world;

## CONCLUSIONS

The economical-financial communication of the companies continues to be characterized by an unsatisfying level of content standardization or by a significant quantity written on paper, which leads to a real discomfort for the economical operators and the national system as an assembly. Surpassing these critical aspects can be eased by enterprising initiatives which to stimulate affirming the reference taxonomy and the technological standards, promoting in the same time investments, aiming to limit the use of paper. The available technology is capable to favor the modernization of financial reporting making, and ease these ones' management both in the definition stage of the structures and also in the change and distribution stage. For this, a first protagonist role, on the international scene, is held by the XBRL standard (eXtensible Business Reporting Language)- a language for the electronic communication of the financial information, lead by an international consortium, present in 18 countries by national jurisdiction, but in which our country does not yet participate.

## REFERENCES

[1] Andrea Fradeani, *La Globalizzazione Della Comunicazione Economico-Finanziaria. IAS/IFRS e XBRL*, Giuffrè 2005, Milan

[2] Deitel, H.M., Deitel, P.J., Nieto, T.R., Lin, T.M., Sadhu, P., *XML: How to Program*, Prentice Hall, 2001;

[3] Gelinas, Ulric J., Jr., Sutton, S.G., Capitolul 14, pp. 515, din *Accounting Information Systems*, 5th edition, South-Western College Pub., 5th edition, 2001;

[4] Hoffman, Ch., *Run XBRL Right Now*, Journal of Accountancy, August, 2000, p.28-29;

[5] Hoffman, Ch., *XBRL Essentials*, American Institute of Certified Public Accountants, 2001;

[6] Luca Erzegovesi, Elena Bonetti, *Utilizzo di XBRL e Quantrix Modeler nelle analisi di bilancio -* Parte 1, Dipartimento di Informatica e Studi Aziendali, Università degli Studi di Trento, 6 febbraio 2007

[7] Wallman, S.M.H., *The Impact of the Internet on the Future of Financial Reporting,* The Brookings Institution

[8] Walter Aste, Davide Panizzolo, ALEA Tech Reports - *Lo standard XBRL (eXtensible Business Reporting Language) e la comunicazione finanziaria d'impresa*, Tech Report numero 20, revisione febbraio 2006 (prima versione: maggio 2004).

[9] http://ifr.SAP.com SAP Business Framework

[10] http://business.fullerton.edu/xbrl Educatia moderna in domeniu

[11] http://xbrlsolutions.com instrumente de lucru XNRL

[12] dev.ekeeper.com, www.kpmg.com crearea taxonomiilor si instrumente de editare

[13] www.msdw.com/xbrl idem

[14] http://www.xbrl.org resurse si soluții

[15] http://www.xbrl.org/demos/demos.htm idem

[16] http://xbrlsolutions.com idem

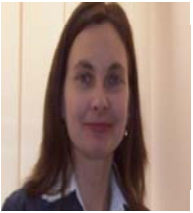
**Veronica Grosu**, **Asistent PhD** - Stefan cel Mare University of Suceava is a co-author of 3 specialty books, over 10 papers published in journals rated by ISI Thompson or within the international symposiums and conferences. She is a member in 3 international professional organizations.

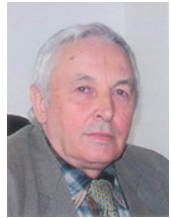
**Rusalim Petris, Prof. PhD** - Stefan cel Mare University, authors 5 speciality books, 35 scientifically papers published in the country and abroad at the International Symposiums or Conferences. He is a member in 10 international professional organizations and scientifically referee in the editing committee of 2 journals rated by REPEC, Socionet, Research Gate etc.

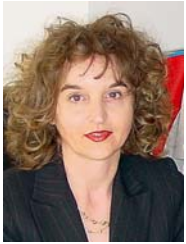
**Elena Hlaciuc -** Stefan cel Mare University, authors 5 speciality books, over 10 papers published in journals rated by ISI Thompson and 30 scientifically papers published in the country and abroad at the International Symposiums or Conferences. He is a member in 10 international professional organizations and scientifically referee in the editing committee of 2 journals rated by REPEC, Socionet, Research Gate etc.

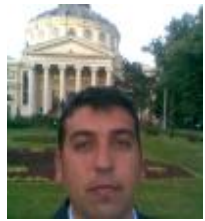
**Marian Socoliuc - Asistent PhD -** Stefan cel Mare University of Suceava is a co-author of 3 specialty books, over 5 papers published in journals rated by ISI Thompson or within the international symposiums and conferences. She is a member in 3 international professional organizations.

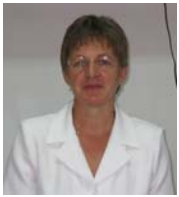
**Eugenia Iancu -** Stefan cel MareUniversity from Suceava. She is a doctorand at Technical Univesity from Timisoara. Authors 3 papers published in Journal rated by ISI Thompson and 25 scientifically papers published in the country and abroad at the International Symposiums or Conferences. She's experienced in research contracts, she's part of the research team in 9 contracted projects, of which 4 are finalized and 5 are current.